\documentclass[aps,pre,twocolumn,eqsecnum,superscriptaddress]{revtex4}    
\usepackage{amsmath}    
\usepackage{amssymb}
\usepackage{subfigure}
\usepackage{epsfig}
\usepackage{array}      

\begin{document}

\title{Shear flow of angular grains: acoustic effects and non-monotonic rate dependence of volume}

\author{Charles K. C. Lieou}
\affiliation{Department of Physics, University of California, Santa Barbara, CA 93106, USA}
\author{Ahmed E. Elbanna}
\affiliation{Department of Civil and Environmental Engineering, University of Illinois, Urbana-Champaign, IL 61801, USA}
\author{J. S. Langer}
\affiliation{Department of Physics, University of California, Santa Barbara, CA 93106, USA}
\author{J. M. Carlson}
\affiliation{Department of Physics, University of California, Santa Barbara, CA 93106, USA}
\date{\today}

\begin{abstract}
Naturally-occurring granular materials often consist of angular particles whose shape and frictional characteristics may have important implications on macroscopic flow rheology. In this paper, we provide a theoretical account for the peculiar phenomenon of auto-acoustic compaction -- non-monotonic variation of shear band volume with shear rate in angular particles -- recently observed in experiments. Our approach is based on the notion that the volume of a granular material is determined by an effective-disorder temperature known as the compactivity. Noise sources in a driven granular material couple its various degrees of freedom and the environment, causing the flow of entropy between them. The grain-scale dynamics is described by the shear-transformation-zone (STZ) theory of granular flow, which accounts for irreversible plastic deformation in terms of localized flow defects whose density is governed by the state of configurational disorder. To model the effects of grain shape and frictional characteristics, we propose an Ising-like internal variable to account for nearest-neighbor grain interlocking and geometric frustration, and interpret the effect of friction as an acoustic noise strength. We show quantitative agreement between experimental measurements and theoretical predictions, and propose additional experiments that provide stringent tests on the new theoretical elements.
\end{abstract}

\maketitle

\section{Introduction}

The purpose of this paper is to explore the peculiar dynamics of a sheared granular material composed of angular grains which are shape-anisotropic and frictional in character. In doing so, we shall provide an explanation of the phenomenon of auto-acoustic compaction, recently observed in a series of experiments by van der Elst \textit{et al.}~\cite{brodsky_2012}, in which the sample volume varies reversibly with the applied shear rate in a non-monotonic fashion. Specifically, their experiments found shear band volume reduction by up to $10\%$ at intermediate shear rates between the slow quasi-static and fast grain-inertial flow regimes for angular sand particles, but not for smooth glass beads, both in the presence and absence of tapping -- forced, periodic vibrational excitation. The authors of that paper posit that shearing provides a source of acoustic energy that un-jams a granular material and allows the granular medium to explore packing configurations. At intermediate shear rates, acoustic vibrations result in a denser packing, similar to compaction due to externally-driven vibrations~\cite{johnson_2008,daniels_2005,daniels_2006,nowak_1998,knight_1995}. Other experiments have also found non-monotonic flow rheology in granular media composed of shape-anisotropic grains~\cite{brodsky_2007,dijksman_2011}. An understanding of the effect of acoustic phenomena in sheared granular flow is especially important in the context of earthquakes, which generate seismic waves that propagate to gouge-filled faults in the vicinity and may cause dramatic reduction in shear strength~\cite{melosh_1979,melosh_1996}.

Our analysis is based primarily on the idea that nonequilibrium states of a granular material are characterized by its compactivity 
\begin{equation}\label{eq:compactivity_def}
 X = \left( \dfrac{\partial V}{\partial S_C} \right)_{\Lambda_{\alpha}} ,
\end{equation}
or equivalently, its effective disorder temperature $T_{\text{eff}} = p X$~\cite{edwards_1989a,edwards_1989b,edwards_1989c,edwards_1990a,edwards_1990b}. (Here $V$ is the extensive volume of the system, $S_C$ is the configurational entropy, and the $\Lambda_{\alpha}$'s are internal variables that specify the configurational state of the granular subsystem.) The state variable $X$ is a thermodynamically well-defined quantity; we naturally assume that the observable volume $V$ is a function of $X$. In this spirit, we assume that $X$ is determined by an entropy-flow equation that is consistent with the second law of thermodynamics. The compactivity $X$ is increased by external work done on the system, and decreases when entropy flows from the granular system into its environment. This heat flow is governed by various noise sources within the system: noise generated by the driving forces, noise generated by friction between particles, etc. Thus, the disorder temperature and the volume are determined by the interplay between these dissipative effects. Specifically, under generic assumptions regarding its functional form, the frictional noise can describe the competition between shear-induced dilation and acoustic compaction, thereby explaining the non-monotonic variation of sample volume with shear rate.

In the present investigation, the microscopic model for effective-temperature dynamics, grain interactions, and the driving forces in the system is the shear-transformation-zone (STZ) theory of granular flow, originally developed to study shear flow in amorphous molecular solids~\cite{falk_1998,langer_2011}. In the context of granular media, the STZ theory has been invoked to account for constitutive friction laws in earthquake physics~\cite{daub_2010}, glassy phenomena in sheared hard-sphere systems~\cite{lieou_2012}, and formation of a finely comminuted gouge layer in fault materials~\cite{lieou_2014}. Under the STZ theoretical framework, plastic deformation can be explained in terms of localized flow defects, or STZ's, whose density is characterized by the compactivity $X$.

Prior applications of the STZ theory made no assumptions regarding the shape and characteristics of constituent grains. The van der Elst \textit{et al.}~experiments, however, clearly demonstrate the significance of grain shape and frictional characteristics in granular flow rheology. The goal of this paper is to provide a quantitative description of these effects. Our proposition is that the large variation of volume is the result of geometric frustration, or lack thereof, between neighboring grains, modeled in terms of an Ising-like internal variable. When coupled with the interpretation of frictional dissipation between particles as a kind of noise, it is possible to quantitatively account for the observed non-monotonic variation of sample volume with shear rate in a granular medium with angular particles.

The rest of our paper is structured as follows. In Sec.~II, we repeat the statistical-thermodynamic analysis largely along the lines of~\cite{lieou_2012,lieou_2014}, but incorporate the effect of tapping and inter-particle frictional dissipation. Specifically, we introduce a ``frictional noise'' which couples the fast, kinetic, and slow, configurational, degrees freedom, and accounts for how frictional dissipation may cause the steady-state sample volume to vary non-monotonically with shear rate. In Sec.~III we introduce our microscopic model that describes how the volume varies with the compactivity. The model is a combination of STZ's and misalignment defects, the latter of which are described by an extra Ising-like internal variable that characterizes the shape effect in terms of grain orientation, interlocking, and geometric frustration. Then, in Sec.~IV we present our theoretical predictions which quantitatively match the experimental measurements of van der Elst \textit{et al}. We conclude the paper in Sec.~V with a list of proposed experiments and future directions.

\section{Theoretical formulation for volume variation}\label{sec:2}

\subsection{Statistical thermodynamics}

As in prior applications of STZ theory~\cite{lieou_2012,lieou_2014}, it is important to quantify the interaction between different components of the granular system; to this end, we turn to the laws of thermodynamics. The developments in this section largely mirror those of~\cite{lieou_2012,lieou_2014}.

Consider a noncrystalline system of hard grains at temperature $T$, with total energy $U_T$. For simplicity we use a single state variable $T$ to characterize all macroscopic and microscopic kinetic-vibrational degrees of freedom of the grains, assumed to be in contact with a thermal reservoir. A consequence of this simplification is that frictional dissipation, among other forms of inelastic grain interaction, simply amounts to the flow of energy from the macroscopic to the microscopic degrees of freedom~\cite{leff_1992,leff_1993}. It is thus unnecessary to account for friction explicitly in the overall energy balance. This simplification is especially convenient in view of our characterization of inter-particle friction as a kind of noise below. For practical purposes, $T$ can be interpreted as a measure of material preparation. Because grains interact only via contact forces, there is no configurational potential energy, so $U_T$ equals the total energy of the system.

Suppose that this system is driven by a shear stress $s$ in the presence of a pressure $p$. The first law of thermodynamics for this system is
\begin{eqnarray}
\label{eq:first_law}
 \nonumber \dot{U}_T &=&  V s\, \dot{\gamma}^{\text{pl}} - p\, \dot{V} \\ &=&  V s\, \dot{\gamma}^{\text{pl}} - p X \dot{S}_C - p \sum_{\alpha} \left( \dfrac{\partial V}{\partial \Lambda_{\alpha}} \right)_{S_C} \dot{\Lambda}_{\alpha}, ~~~~~
\end{eqnarray}
where $\dot{\gamma}^{\text{pl}}$ is the plastic shear rate. As in Eq.~(\ref{eq:compactivity_def}), $S_C$ is the granular configurational entropy, and the $\Lambda_{\alpha}$ are internal variables that specify the configurational state of the granular subsystem.  

Let $S_T$ denote the entropy of all kinetic-vibrational degrees of freedom.  Then
\begin{equation}
 \dot{U}_T = T \dot{S}_T ,
\end{equation}
and 
\begin{equation}\label{eq:S_C}
 p X \dot{S}_C = V s \dot{\gamma}^{\text{pl}} - p \sum_{\alpha} \left( \dfrac{\partial V}{\partial \Lambda_{\alpha}} \right)_{S_C} \dot{\Lambda}_{\alpha} - T \dot{S}_T .
\end{equation}
The second law of thermodynamics states that the total entropy of the system, being the sum of all kinetic-vibrational and configurational degrees of freedom, must be a non-decreasing function of time:
\begin{equation}\label{eq:second_law}
 \dot{S} = \dot{S}_C + \dot{S}_T \geq 0 .
\end{equation}
Substituting Eq.~\eqref{eq:S_C} for $\dot S_C$ into the second law above, and using the fact that each individually variable term in the resulting inequality must be non-negative \cite{coleman_1963,langer_2009a,langer_2009b,langer_2009c}, we arrive at the second-law constraints
\begin{eqnarray}
 \label{eq:W} {\cal W} = V s\, \dot{\gamma}^{\text{pl}} - p \sum_{\alpha} \left( \dfrac{\partial V}{\partial \Lambda_{\alpha}} \right)_{S_C} \dot{\Lambda}_{\alpha} \geq 0; \\
 \label{eq:S_T} (p X - T) \dot{S}_T \geq 0 .
\end{eqnarray}
In arriving at these two constraints, we have arranged terms in such a way that terms pertaining to the degrees of freedom that belong to the same subsystem are grouped together. The dissipation rate ${\cal W}$, as defined in \cite{langer_2009a,langer_2009b,langer_2009c}, is the difference between the rate at which inelastic work is done on the configurational subsystem and the rate at which energy is stored in the internal degrees of freedom. The second constraint implies that $p X - T$ and $\dot{S}_T$ must carry the same sign if they are nonzero, so that
\begin{equation}\label{eq:Q}
 T \dot{S}_T = - {\cal K} (X, T) \left( T - p\, X \right)\equiv \,{\cal Q},
\end{equation}
where ${\cal K} (X, T)$ is a non-negative thermal transport coefficient. It is already clear from this analysis that $p\,X$ plays the role of a temperature. $p\,X$ approaches $T$ in an equilibrating system; and a heat flux ${\cal Q}$ flows between the granular subsystem and the reservoir when the two subsystems are not in thermodynamic equilibrium with each other.

\subsection{Steady-state compactivity as a function of strain rate}

Let us introduce the dimensionless strain rate, or the inertial number (see for example~\cite{jop_2006})
\begin{equation}
 q \equiv \tau \dot{\gamma}^{\text{pl}} ,
\end{equation}
where $\tau$ is the inertial time scale, to be discussed in greater detail in Sec.~III. (In the steady state, the total, imposed shear rate $\dot{\gamma}$ equals the plastic strain rate $\dot{\gamma}^{\text{pl}}$.) We also define the dimensionless compactivity
\begin{equation}
 \chi \equiv X / v_Z ,
\end{equation}
where $v_Z$ is the excess volume associated with STZ's -- rare, non-interacting loose spots where irreversible particle rearrangements occur. (In the solidlike -- as opposed to hydrodynamic -- regime, nonaffine particle displacements occur everywhere. However, 1STZ's refer to the subset of those that are irreversible and involve local topological change, so our picture of dilute defects remains valid.) The dimensionless compactivity $\chi$ measures the amount of configurational (i.e.~ structural) disorder in the granular system. Intuitively, $\chi$ increases monotonically with the extensive volume $V$ of the system. It is obvious that external vibrations and inter-particle dissipative mechanisms such as friction play important roles in controlling the configurational state of the granular medium. However, in the absence of these mechanisms, as in hard-sphere systems with zero vibrational noise strength~\cite{lieou_2012}, the steady-state compactivity ought to be a function of the strain rate alone: $\chi = \hat{\chi} (q)$. As seen in our hard-sphere analysis and in simulations~\cite{lieou_2012,liu_2011,haxton_2012}, $\hat{\chi} (q)$ approaches some constant $\hat{\chi}_0$ in the limit of small $q$. On the other hand, $\hat{\chi} (q)$ becomes a rapidly increasing function of $q$ once the shear rate becomes comparable to the rate of intrinsic structural relaxation, to reflect shear-rate dilation in hard-sphere systems. It is customary to write the inverse relation $q (\hat{\chi})$ in the Vogel-Fulcher-Tamann (VFT) form~\cite{lieou_2012,haxton_2007,manning_2007b}
\begin{equation}\label{eq:vft}
 \dfrac{1}{q} = \dfrac{1}{q_0} \exp \left[ \dfrac{A}{\hat{\chi}} + \alpha_{\text{eff}} (\hat{\chi}) \right] ,
\end{equation}
where
\begin{equation}
 \alpha_{\text{eff}} (\hat{\chi}) = \left( \dfrac{\hat{\chi}_1}{\hat{\chi} - \hat{\chi}_0} \right) \exp \left( - 3 \dfrac{\hat{\chi} - \hat{\chi}_0}{\hat{\chi}_A - \hat{\chi}_0} \right) .
\end{equation}

The quantity $\chi$ evolves according to the first law of thermodynamics. To deduce its equation of motion, return to Eq.~\eqref{eq:S_C} for the rate of entropy change $\dot{S}_C$ of the configurational subsystem, let us invoke the quasistationary approximation $\dot{\Lambda}_{\alpha} = 0$ for each internal variable $\Lambda_{\alpha}$, and use Eq.~\eqref{eq:Q} to eliminate $\dot{S}_T$. The result is
\begin{equation}\label{eq:S_C2}
 p X \dot{S}_C = V s \dot{\gamma}^{\text{pl}} - {\cal K} (X, T) (p X - T) .
\end{equation}
To convert this into an equation for the dimensionless compactivity $\chi$, we use the scaling $\chi = X / v_Z$ and $\theta \equiv T / p \, v_Z$. Then, we use the relation
\begin{eqnarray}\label{eq:chiscdot}
 \nonumber \chi \dot{S}_C &=& \chi \left( \dfrac{\partial S_C}{\partial \chi} \right)_{\Lambda_{\alpha}} \dot{\chi} + \chi \sum_{\alpha} \left( \dfrac{\partial S_C}{\partial \Lambda_{\alpha}} \right)_{S_C} \dot{\Lambda}_{\alpha} \\ &=& \chi \left( \dfrac{\partial S_C}{\partial \chi} \right)_{\Lambda_{\alpha}} \dot{\chi},
\end{eqnarray}
where in the second equality we again used the quasistationary approximation to eliminate the time derivatives of other internal variables. Now comes the crucial step: since the transport coefficient ${\cal K} (\chi, \theta)$ couples the configurational and kinetic-vibrational subsystems, it should consist of additive mechanical, vibrational and frictional contributions. Specifically,
\begin{equation}\label{eq:A}
 {\cal K} (\chi, \theta) = \dfrac{V}{\tau} \, {\cal A} \, \left( \Gamma + \xi + \rho \right) .
\end{equation}
Here, $\Gamma$ is the mechanical noise that pertains to externally applied shear. It will be computed below in Eq.~\eqref{eq:Gamma} in terms of the rate of entropy generation, and we will show that it is proportional to the work of plastic deformation, i.e., the tensor product $s \dot{\gamma}^{\text{pl}}$ of the shear stress and the plastic strain rate. On the other hand, the dimensionless quantity $\rho$ is a measure of the intensity of externally-imposed acoustic-vibrational motion of the grains~\cite{brodsky_2012}, more generally known as tapping. Tapping provides a means to un-jam a granular system so that it can explore packing configurations. In granular experiments, acoustic vibrations have been found to increase the packing fraction~\cite{daniels_2005,daniels_2006,nowak_1998,knight_1995}, and trigger stick-slip motion under shear~\cite{johnson_2008}. When $\rho = 0$, the system is fully jammed in the sense that configurational rearrangements can occur only in response to sufficiently large driving forces. In addition, $\xi$ is the system-specific frictional coupling or noise, to be determined based on phenomenology. In contrast to prior STZ analyses in which no assumption whatsoever was made in regard to the dissipative nature of particle interaction~\cite{lieou_2012,langer_2009c,langer_2008}, the $\rho$ term is replaced by $\xi + \rho$ to reflect the importance of frictional dissipation. ${\cal A}$ is a non-negative quantity to be determined by appealing to the steady-state solution in special cases. We also implicitly subsume all time scales relevant to tapping and friction under the inertial time scale $\tau$ in Eq.~\eqref{eq:A}.

In anticipation of Pechenik's hypothesis in Eq.~\eqref{eq:Gamma} below, we rewrite the first term in Eq.~\eqref{eq:S_C2}: $s \dot{\gamma}^{\text{pl}} = ( \Gamma / \tau) {\cal B}$, where ${\cal B}$ is a constant. Then, some simple algebra, along with use of Eqs.~\eqref{eq:chiscdot} and~\eqref{eq:A}, reduce Eq.~\eqref{eq:S_C2} to
\begin{equation}
 c^{\text{eff}} \dot{\chi} = \Gamma {\cal B} - {\cal A} \left( \Gamma + \xi + \rho \right) (\chi - \theta),
\end{equation}
with $c^{\text{eff}}$ being a scalar quantity that describes the capacity of volume dilation. We argued above that in the absence of vibration or inelastic dissipation, $\rho = \xi = 0$, the steady-state compactivity is uniquely determined as $\chi^{\text{ss}} = \hat{\chi} (q)$. This gives ${\cal A} = {\cal B} / (\hat{\chi} (q) - \theta )$ whose non-negativity incidentally implies the constraint that $\hat{\chi} (q) > \theta$, i.e., stirring the system drives 
the slow, configurational degrees of freedom out of equilibrium with the fast, kinetic-vibrational degrees of freedom. Then, in general, the steady-state compactivity is given by
\begin{equation}\label{eq:X_ss}
 \chi^{\text{ss}} = \dfrac{\Gamma \hat{\chi}(q) + (\xi + \rho) \theta}{\Gamma + (\xi + \rho)} .
\end{equation}

As we alluded to above, the extensive volume $V$ of the system -- measured in~\cite{brodsky_2012} by the change in shear band thickness -- is an increasing function of the compactivity $\chi$. The explicit functional form of $V(\chi)$ will be discussed later in Sec.~III in the context of a microscopic model and internal state variables. Thus Eq.~\eqref{eq:X_ss}, in effect, describes the variation of system volume with shear rate and noise strength.

With $\hat{\chi}(q)$ being an increasing function of the dimensionless strain rate $q$, how can we understand the non-monotonic variation of shear band thickness -- and therefore the compactivity $\chi$ -- with shear rate, as observed in the experiments of van der Elst \textit{et al.}~\cite{brodsky_2012}, within Eq.~\eqref{eq:X_ss}? Specifically, can we account for the decrease in $\chi$ at intermediate strain rates, and shear-rate dilation at large $q$? The answer lies in the frictional noise term $\xi$. Intuitively, $\xi$ should be a scalar function of the plastic work of shearing; thus $\xi = \xi (\Gamma)$. It induces correlations between particle velocities and enhances non-local effects~\cite{kamrin_2014}. Let us first focus on the case when vibrations are absent. In the limit of vanishingly small strain rate $q$, the mechanical noise $\Gamma \rightarrow 0$; Eq.~\eqref{eq:X_ss} then shows that $\chi^{\text{ss}} \rightarrow \theta$ provided that $\xi \neq 0$. Because $\hat{\chi} (q)$ increases monotonically with $q$ and exceeds $\theta$, $\chi^{\text{ss}}$ must also be an increasing function of $q$, which is contrary to the non-monotonicity of volume variation with shear rate. A resolution to this dilemma is that $\xi (\Gamma) = 0$ at zero shear rate, to reflect the fact that friction does not dissipate energy when no slipping takes place. In fact, if $\xi (\Gamma) \rightarrow 0$ faster than $\Gamma$ at small shear rates -- say, if $\xi (\Gamma) \sim \Gamma^2$ for small $\Gamma$, as in Newtonian friction -- then $\chi^{\text{ss}} \rightarrow \hat{\chi} (q = 0) = \hat{\chi}_0$ in that limit. (Indeed, if we interpret $\xi$ as some kind of energy, in analogy to $\Gamma$, then because the energy associated with an inelastic collision between grains is proportional to the square of their relative velocity, it is plausible for $\xi (\Gamma) \sim \Gamma^2$.) As $q$ increases so that $\xi (\Gamma)$ becomes large enough, it is possible for the steady-state compactivity $\chi^{\text{ss}}$ to fall below $\hat{\chi} (q)$: $\theta < \chi^{\text{ss}} < \hat{\chi} (q)$, because $\hat{\chi} (q) > \theta$.

In the opposite limit of large shear rate, the experiments indicate that shear-induced dilation must once again dominate, and that inter-particle friction becomes less important. Thus we stipulate that $\xi (\Gamma)$ saturates at large $\Gamma$ so that $\chi^{\text{ss}} \rightarrow \hat{\chi} (q)$ in Eq.~\eqref{eq:X_ss}. We now check that this assumption is consistent with experimental findings in the presence of tapping. For small $q$, Eq.~\eqref{eq:X_ss} with $\rho \neq 0$ indicates that $\chi^{\text{ss}} \rightarrow \theta < \hat{\chi}_0$, so that tapping does increase the packing fraction in the slow quasistatic limit. On the other hand, the boundedness of both $\xi$ and $\rho$ shows that $\chi^{\text{ss}} \rightarrow \hat{\chi}(q)$ in the fast inertial regime, independent of friction and tapping, and coinciding with the $\rho = 0$ behavior as seen by the overlap of the two curves in Fig.~\ref{fig:Vq} below in that limit.

Having argued that non-monotonic variation of volume with shear rate is indeed possible, let us now turn our attention to formulating a microscopic model that accounts for the flow rheology of angular grains, and quantifies volume variation with configurational disorder.

\section{Microscopic model}

\subsection{Internal variables and system volume}

The extensive volume $V$ of the granular packing plays a central role in this paper. As such, we characterize the volume in terms of the configurational state of the system, and derive equations of motion for the corresponding internal variables. We emphasize that the non-monotonic variation of volume with shear rate as a result of inter-particle friction is not unique to the microscopic model to be introduced here. In fact, any model for which the volume $V$ varies monotonically with the compactivity $X$ ought to qualitatively describe this non-monotonicity. However, the value of our model lies in its ability to provide a physical account of the relationship between grain-scale configuration and volume, as well as a good quantitative fit to the experimental data. Readers who are not interested in the microscopic details may skip directly to Sec.~III.~E, where we summarize the formulas that will be used in the ensuing analysis.

Recall our physical picture that in dense granular flow, irreversible particle rearrangements occur at rare, non-interacting soft spots with excess free volume known as STZ's. The applied shear stress defines a direction relative to which STZ's can be classified according to orientation, with total numbers $N_+$ and $N_-$ respectively. Upon application of shear stress in the ``plus'' direction, STZ's of the minus type easily deform to become plus-type STZ's. However, plus-type STZ's rarely flip and acquire the minus orientation; rather, they are annihilated readily by noise. If the total number of grains equals $N$, we define the intensive variables
\begin{equation}
 \Lambda = \dfrac{N_+ + N_-}{N}; \quad m = \dfrac{N_+ - N_-}{N_+ + N_-}
\end{equation}
as the density and orientational bias of STZ's.

On the other hand, in angular grains, shape anisotropy and geometric frustration allows for the distinct possibility for grains interlocking, which reduces local volume, independent of the presence of localized slip events. Said differently, because of the absence of infinite-fold symmetry, neighboring grains that do not align with one another contribute excess volume. The simplest way to describe this is to assume that there is an extra Ising-like order parameter, $\eta$, that describes grain orientation (this orientation is independent of the direction of shear stress). Specifically, let $N^G_+$ and $N^G_-$ denote the number of grains in each of the two orientations, and define
\begin{equation}\label{eq:eta_def}
 \eta = \dfrac{N^G_+ - N^G_-}{N} .
\end{equation}
Of course, $-1 \leq \eta \leq 1$, as it should. Unlike STZ's which are rare, isolated defects, each grain is associated with a particular direction; that is, $N^G_+ + N^G_- = N$.

Denote by $v_Z$ and $v_a$ the excess volumes associated with STZ's and misalignments. The assumption that the system volume does not depend on STZ orientation, but depends on nearest-neighbor interactions in an Ising-like manner similar to the binary clusters recently invoked in a model of glass transition~\cite{langer_2013}, allows us to write down the extensive volume of the system as follows:
\begin{eqnarray}\label{eq:V}
 \nonumber V &=& V_0 + N \Lambda v_Z - N \eta^2 v_a + V_1 (S_1) \\\nonumber &=& V_0 + N \Lambda v_Z - N \eta^2 v_a \\ & & + V_1 (S_C - S_Z (\Lambda, m) - S_G (\eta) ). ~~~~~
\end{eqnarray}
Here, $V_0 = N a^3$ is the total volume of grains. $S_C$ is the configurational entropy, consisting of the entropy $S_Z$ associated with STZ's, $S_G$ associated with grain orientation, and $S_1$ for all other configurational degrees of freedom. Correspondingly, $V_1$ is the volume associated with those degrees of freedom, to be discussed in greater detail towards the end of this Section. Then, under the assumption the STZ's and grain alignments are two-state entities, we can compute $S_Z$ and $S_G$ easily by counting the number of possible configurations of distributing $N_+$ and $N_-$ STZ's of each orientation, and $N^G_+$ and $N^G_-$ orientation states for each grain, among $N$ sites~\cite{langer_2009c}. The result is
\begin{eqnarray}
 \label{eq:S_Z} S_Z (\Lambda, m) &=& N S_0 (\Lambda) + N \Lambda \psi (m);\\
 \label{eq:S_G} S_G (\eta) &=& N \psi (\eta) ,
\end{eqnarray}
where
\begin{eqnarray}
 \label{eq:S_0} S_0 (\Lambda) &=& - \Lambda \ln \Lambda + \Lambda;\\
 \label{eq:psi}\nonumber \psi(m) &=& \ln 2 - \dfrac{1}{2} (1 + m) \ln (1 + m) \\ & & - \dfrac{1}{2} (1 - m) \ln (1 - m).
\end{eqnarray}

\subsection{Equations of motion}

In order to study the dynamics of the system, the preceding analysis needs to be supplemented with equations of motion for each of the internal variables. We first look at STZ dynamics. As usual, the STZ equation of motion for $N_+$ and $N_-$ is given by:
\begin{equation}\label{eq:master}
 \tau \dot{N}_{\pm} = {\cal R} (\pm s) N_{\mp} - {\cal R} (\mp s) N_{\pm} + \tilde{\Gamma} \left( \dfrac{1}{2} N^{\text{eq}} - N_{\pm} \right) .
\end{equation}
The corresponding strain rate is
\begin{equation}\label{eq:strainrate}
 \dot{\gamma}^{\text{pl}} = \dfrac{2\,v_0}{\tau V} \left[ {\cal R}(s) N_- - {\cal R}(-s) N_+ \right],
\end{equation}
where we define the volume of the plastic core of an STZ to be $v_0$, which ought to be proportional to $a^3$. Because we are describing simple rather than pure shear, there is a factor of $2$ up front.

Some comments on the various quantities that appear in Eqs.~\eqref{eq:master} and~\eqref{eq:strainrate} are now in order. The quantity $\tau = a \sqrt{\rho_G / p}$, where $\rho_G$ denotes the material density of the grains, is the inertial time scale that characterizes the typical duration of a pressure-driven particle rearrangement event~\cite{lieou_2012,lieou_2014,liu_2011,haxton_2012}. It is proportional to the average time between successive grain-grain collisions. This time scale also applies in a dense granular medium where inter-particle friction is important, as long as the friction is proportional to the normal force at the contact interface. Its product with the shear rate $\dot{\gamma}^{\text{pl}}$ gives the so-called inertial number, the magnitude of which determines the flow regime of dense granular flow~\cite{jop_2006}. ${\cal R}(\pm s)$ represent the rates (in units of $\tau^{-1}$) at which the STZ's are making forward and backward transitions. The term proportional to $\tilde{\Gamma}$ represents the rates of STZ creation and annihilation; $N^{\text{eq}}$ is the steady-state, total number of STZ's. Specifically, $\tilde{\Gamma}/\tau$ is an attempt frequency consisting of additive vibrational and mechanical parts: 
\begin{equation}
\tilde{\Gamma} = \rho +\Gamma.
\end{equation}
Because friction should play no role in the creation or annihilation of STZ's, $\xi$ does not appear in the expression for $\tilde{\Gamma} / \tau$.

After making these remarks on the elements of the extensive STZ equations of motion, we can rewrite them exclusively in terms of the intensive state variables $\Lambda$ and $m$ as follows:
\begin{eqnarray}
 \label{eq:Lambda} \tau\, \dot{\Lambda} &=& \tilde{\Gamma} ( \Lambda^{\text{eq}} - \Lambda ) ; \\
 \label{eq:m} \tau\, \dot{m} &=& 2\, {\cal C}(s) ( {\cal T}(s) - m ) - \tilde{\Gamma} m - \tau \dfrac{\dot{\Lambda}}{\Lambda} m ; ~~~~~ \\
 \label{eq:D_pl} \tau \,\dot{\gamma}^{\text{pl}} &=&2\, \epsilon_0\,\Lambda\, {\cal C}(s) ({\cal T}(s) - m),
\end{eqnarray}
where $\epsilon_0 = N\, v_0 / V_0$ is independent of $a$, and $\Lambda^{\text{eq}} = N^{eq} / N$. In writing Eq.~\eqref{eq:D_pl} we have implicitly made the approximation $\Lambda \ll 1$ and $V_1 \ll V_0$ in Eq.~\eqref{eq:V} so that $V \approx V_0$ in Eq.~\eqref{eq:strainrate}. We also define
\begin{equation}
\label{eq:calC}
 {\cal C}(s) = \dfrac{1}{2} \left( {\cal R}(s) + {\cal R}(-s) \right) ,
\end{equation}
and
\begin{equation}
 {\cal T}(s) = \dfrac{{\cal R}(s) - {\cal R}(-s)}{{\cal R}(s) + {\cal R}(-s)} .
\end{equation}

In analogy to the master equation for STZ transitions, we propose that the simplest possible equation of motion that describes the change in the number of grains is of the form
\begin{equation}\label{eq:NG}
 \tau \dot{N}_{\pm}^G = {\cal R}_{\pm}^G N_{\mp}^G - {\cal R}_{\mp}^G N_{\mp}^G .
\end{equation}
Here, ${\cal R}_{\pm}^G$ is a yet-to-be specified rate factor for the transition between orientations, absorbing all other relevant time scales such as the inverse tapping frequency; we expect that it is proportional to the sum of mechanical and vibrational noise strengths $\tilde{\Gamma}$. A key difference between Eq.~\eqref{eq:NG} and its counterpart, Eq.~\eqref{eq:master} for STZ transitions, is the absence of an extra term which, in Eq.~\eqref{eq:master}, describes the creation and annihilation of STZ's and the approach of STZ density to an equilibrium value. The reason behind this is two-fold. Firstly, as we alluded to before, every grain must belong to either one of the two orientations, but a given grain need not be part of an STZ at a given instant. Secondly, the effect of noise is already accounted for in the rate factor ${\cal R}_{\pm}^G$ which is not directly related to the direction of the shear stress $s$. With this, the STZ equations of motion is supplemented by an extra equation for the temporal evolution of grain orientational bias $\eta$:
\begin{equation}\label{eq:eta}
 \tau \dot{\eta} = 2 {\cal C}^G ({\cal T}^G - \eta),
\end{equation}
where
\begin{equation}
 {\cal C}^G = \dfrac{1}{2} ( {\cal R}^G_+ + {\cal R}^G_- ); \quad {\cal T}^G = \dfrac{{\cal R}^G_+ - {\cal R}^G_-}{{\cal R}^G_+ + {\cal R}^G_-} .
\end{equation}

\subsection{Dissipation rate and thermodynamic constraints}

At this point in the development, the second law of thermodynamics provides useful constraints on various ingredients of the equations of motion, and on steady-state dynamics of the system. To this end, we now substitute this and Eqs.~\eqref{eq:Lambda}, \eqref{eq:m}, \eqref{eq:D_pl}, and \eqref{eq:eta} into Eq.~\eqref{eq:W} for the dissipation rate ${\cal W}$ which, according to the second law of thermodynamics, must be non-negative. We also use the approximation $V \approx V_0$ where appropriate. The result is
\begin{eqnarray}\label{eq:W2}
 \nonumber \tau\, \dfrac{{\cal W}}{N} &=& - \tilde{\Gamma}\, p\, X\, \Lambda\, m \dfrac{d \psi}{dm} - p\, \tilde{\Gamma}\,(\Lambda^{\text{eq}} - \Lambda) \\\nonumber && \times \left[ v_Z + X \left( \ln \Lambda - \psi (m) + m \,\dfrac{d \psi}{dm} \right) \right] \\\nonumber && + 2\, \Lambda\, {\cal C}(s)\,\Bigl( {\cal T}(s) - m\Bigr) \left( v_0 s + p X \dfrac{d \psi}{dm} \right) \\ && + 2\, {\cal C}^G \, \Bigl( {\cal T}^G - \eta \Bigr) \left( X \dfrac{d \psi}{d \eta} + 2\, \eta \, v_a \right) . ~~~~~
\end{eqnarray}
The second-law constraint, ${\cal W} \geq 0$, must be satisfied by all possible motions of the system; this is guaranteed if each of the four terms in Eq.~\eqref{eq:W2} is non-negative~\cite{coleman_1963,langer_2009a,langer_2009b,langer_2009c}. (Indeed, only the third term depends explicitly on the shear stress $s$, while the second term is proportional to $\dot{\Lambda}$, and $\eta$ appears only in the fourth term. The entire expression must be non-negative irrespective of $s$ and $\dot{\Lambda}$, and $\eta$.) The first term automatically satisfies this requirement because, from Eq.~\eqref{eq:psi}, we have
\begin{equation}\label{eq:dpsidm}
 \dfrac{d \psi}{dm} = - \dfrac{1}{2} \ln \left( \dfrac{1 + m}{1 - m} \right) = - \tanh^{-1} (m) ,
\end{equation}
so that the product $- m ( d \psi / dm)$ is automatically non-negative. 

The non-negativity constraint on the second term in Eq.~\eqref{eq:W2} can be written in the form
\begin{equation}
 - \dfrac{\partial F}{\partial \Lambda} (\Lambda^{\text{eq}} - \Lambda) \geq 0 ,
\end{equation}
where $F$ is a free energy given by
\begin{equation}
 F (\Lambda, m) = p \left[ v_Z \Lambda - X S_0 (\Lambda) - X \Lambda \left( \psi(m) - m \dfrac{d \psi}{d m} \right) \right].
\end{equation}
$\Lambda^{\text{eq}}$ must be the value of $\Lambda$ at which $ \partial F / \partial \Lambda$ changes sign, so that
\begin{equation}\label{eq:Lambda_eq}
 \Lambda^{\text{eq}} = \exp \left[ - \dfrac{v_Z}{X} + \psi(m) - m \dfrac{d \psi}{dm} \right] \approx 2\, \exp \left( - \dfrac{v_Z}{X} \right). ~~~~~
\end{equation}
Thus, the STZ density in this non-equilibrium situation is given by a Boltzmann-like expression in which the compactivity plays the role of the temperature.

As for the third term, we have
\begin{equation}\label{eq:Ts}
 \Bigl( {\cal T}(s) - m \Bigr) \left( v_0 s + p X \dfrac{d \psi}{dm} \right) \geq 0.
\end{equation}
The two factors on the left-hand side must be monotonically increasing functions of $s$ that change sign at the same point for arbitrary values of $m$. According to Eq.~\eqref{eq:dpsidm}, this is possible only if
\begin{equation}
 {\cal T}(s) = \tanh \left( \dfrac{v_0 s}{p X} \right).
\end{equation}

Finally, the extra constraint associated with the new internal variable $\eta$ is
\begin{equation}
 2 {\cal C}^G ({\cal T}^G - \eta) \left( X \dfrac{d \psi}{d \eta} + 2 \eta v_a \right) \geq 0.
\end{equation}
In the same spirit as for the third term, this holds if and only if each of the two quantities in the pair of parenthesis change sign at the same value of $\eta$; thus
\begin{equation}
 {\cal T}^G = \tanh \left( \dfrac{2 \eta v_a}{X} \right) .
\end{equation}
According to Eq.~\eqref{eq:eta}, the steady-state value of $\eta$ is then given by the solution to the equation
\begin{equation}\label{eq:eta_eq}
 \eta^{\text{eq}} = \tanh \left( \dfrac{2 \eta^{\text{eq}} v_a}{X} \right).
\end{equation}
This is reminiscent of the familiar spontaneous symmetry breaking in the magnetization of an Ising ferromagnet. If $X > X_c \equiv 2 v_a$, then the only solution to Eq.~\eqref{eq:eta_eq} is $\eta^{\text{eq}} = 0$; this applies to a ``dilute'' granular packing, for which interlocking is no longer important. On the other hand, if $X < X_c$, there are two non-zero solution in the steady state, $\eta^{\text{eq}} = \pm \eta_0$, the absolute value of which decreases with increasing $X$ in this regime.

\subsection{Mechanical noise and steady-state dynamics}

In quasistationary or steady-state situations such as the experiments by van der Elst \textit{et al.}~\cite{brodsky_2012} that we analyze in this paper, a major simplification comes from setting $\dot{\Lambda} = \dot{m} = \dot{\eta} = 0$, so that $\Lambda = \Lambda^{\text{eq}}$ and $\eta = \eta^{\text{eq}}$. To determine the mechanical noise strength $\Gamma$ and the stationary value of $m$, we invoke Pechenik's hypothesis~\cite{langer_2009c,pechenik_2003}, which states that the mechanical noise strength $\Gamma$ is proportional to the mechanical work per STZ. The plastic work per unit volume is simply $\dot{\gamma}^{\text{pl}} s$. To convert this rate into a noise strength with dimensions of inverse time, we multiply by the volume per STZ, $V_0 / (N \Lambda^{\text{eq}})$, and divide by an energy conveniently written in the form $\epsilon_0 (V_0 / N)\, s_0$. Here, $s_0$ is a system-specific parameter with the dimensions of stress. The resulting expression for $\Gamma$ is
\begin{equation}\label{eq:Gamma}
 \Gamma = \dfrac{\tau \dot{\gamma}^{\text{pl}} s}{\epsilon_0 s_0 \Lambda^{\text{eq}}} = \dfrac{2 s}{s_0} {\cal C}(s) \Bigl( {\cal T}(s) - m \Bigr).
\end{equation}
With this result, the stationary version of Eq.~\eqref{eq:m} reads
\begin{equation}
 2\, {\cal C}(s) \Bigl( {\cal T}(s) - m \Bigr) \left( 1 - \dfrac{m s}{s_0} \right) - m\, \rho  = 0 .
\end{equation}
The stationary value of $m$ is then given by
\begin{eqnarray}\label{eq:m_eq}
 \nonumber m^{\text{eq}} (s) &=& \dfrac{s_0}{2s} \left[ 1 + \dfrac{s}{s_0} {\cal T}(s) + \dfrac{\rho}{2 {\cal C}(s)} \right] \\ &&- \dfrac{s_0}{2s} \sqrt{\left[ 1 + \dfrac{s}{s_0} {\cal T}(s) + \dfrac{\rho}{2 {\cal C}(s)} \right]^2 - 4 \dfrac{s}{s_0} {\cal T}(s)} . ~~~~~
\end{eqnarray}
In particular, when $\rho = 0$, we have
\begin{equation}
 m^{\text{eq}} = 
\begin{cases}
{\cal T}(s), & \text{if $(s / s_0)\, {\cal T}(s) < 1$} ; \\
s_0/s, & \text{if $(s / s_0)\, {\cal T}(s) \geq 1$}.
\end{cases}
\end{equation}
An important consequence of this is that the yield stress for a completely jammed system is the solution of the equation
\begin{equation}
 s_y \, {\cal T}(s_y) = s_y \tanh \left( \dfrac{v_0 s_y}{p X}\right) = s_0.
\end{equation}
If the temperature-like quantity $p\,X$ is small in comparison with $\epsilon_0 a^3 s_0$, then $s_y \approx s_0$. Thus $s_0$ sets, in effect, the minimum flow stress of the system in the absence of tapping. When $\rho \neq 0$, however, the system is un-jammed, and flows at arbitrarily small shear stress $s$.

Finally, let us return to using dimensionless variables $q = \tau \dot{\gamma}^{\text{pl}}$ and $\chi = X / v_Z$. The steady-state version of Eq.~\eqref{eq:D_pl} for the strain rate becomes
\begin{equation}\label{eq:q_s}
 q \equiv \tau \dot{\gamma}^{\text{pl}} = 4\,\epsilon_0\,e^{-\,1/\chi}\,{\cal C}(s) \left[ \tanh \left( \dfrac{\epsilon_0 s}{\epsilon_Z p \chi} \right) - m^{\text{eq}} (s) \right] ,
\end{equation}
where $\epsilon_Z \equiv v_Z / a^3$. Now, from Eq.~\eqref{eq:V} above, the system volume varies monotonically with the compactivity as follows:
\begin{equation}\label{eq:V2}
 \dfrac{V}{V_0} = 1 + \Lambda^{\text{eq}} \epsilon_Z - (\eta^{\text{eq}})^2 \epsilon_a + \dfrac{V_1}{V_0} ,
\end{equation}
where $\epsilon_a = v_a / a^3$, $\Lambda^{\text{eq}} = 2 e^{-1 / \chi^{\text{ss}} }$, and $\eta^{\text{eq}}$ satisfies $\eta^{\text{eq}} = \tanh (2 \eta^{\text{eq}} \epsilon_a / \epsilon_Z \chi^{\text{ss}} )$. We now specify the volume $V_1$ associated with all other configurational degrees of freedom. Because STZ's are rare density fluctuations whose existence results in denser spots nearby, their contribution to the system volume should be small in comparison to the effects of grain interlocking and the packing fraction itself, the latter being subsumed in $V_1$. The simplest assumption is that this volume varies linearly with the compactivity $\chi$: $\Lambda^{\text{eq}} \epsilon_Z + V_1 / V_0 \approx \epsilon_1 (\chi - \chi_r)$, where the effective volume expansion coefficient $\epsilon_1$ is assumed to be a constant, and $\chi_r$ is an offset that can be conveniently chosen to equal $\hat{\chi}_0$. Thus 
\begin{equation}\label{eq:V3}
 \dfrac{V}{V_0} = 1 - (\eta^{\text{eq}})^2 \epsilon_a + \epsilon_1 (\chi - \hat{\chi}_0) .
\end{equation}

\subsection{Summary: steady-state relations between volume, compactivity and shear rate}

Summarizing, the steady-state system volume $V$ varies with the compactivity $\chi^{\text{ss}}$ according to Eq.~\eqref{eq:V3}:
\begin{equation}\label{eq:V4}
 \dfrac{V}{V_0} = 1 - (\eta^{\text{eq}})^2 \epsilon_a + \epsilon_1 (\chi^{\text{ss}} - \hat{\chi}_0) ,
\end{equation}
where $\eta^{\text{eq}}$ satisfies $\eta^{\text{eq}} = \tanh (2 \eta^{\text{eq}} \epsilon_a / \epsilon_Z \chi^{\text{ss}} )$. On the other hand, the steady-state compactivity is controlled by the driving forces and the frictional noise according to Eq.~\eqref{eq:X_ss}:
\begin{equation}\label{eq:X_ss2}
 \chi^{\text{ss}} = \dfrac{\Gamma \hat{\chi}(q) + (\xi + \rho) \theta}{\Gamma + (\xi + \rho)} .
\end{equation}
Here, the dimensionless strain rate is related to the compactivity $\chi^{\text{ss}}$ by
\begin{equation}\label{eq:q_s2}
 q = 4\,\epsilon_0\,e^{-\,1/\chi^{\text{ss}} }\,{\cal C}(s) \left[ \tanh \left( \dfrac{\epsilon_0 s}{\epsilon_Z p \chi} \right) - m^{\text{eq}} (s) \right] ,
\end{equation}
with $m^{\text{eq}}(s)$ given by Eq.~\eqref{eq:m_eq}. Eqs.~\eqref{eq:V4},~\eqref{eq:X_ss2} and~\eqref{eq:q_s2} are the primary relations that describe the variation of system volume, or shear band thickness, with the imposed shear rate $q$.

\section{Analysis of experiments by van der Elst et al.}

Figure 1 shows the rescaled experimental data of van der Elst \textit{et al.}~\cite{brodsky_2012} along with the corresponding results for the theoretical model presented in this paper. The data points represent averages over repeated experimental measurements, and the error bars represent the corresponding standard deviation. The curves show the corresponding results for our theoretical model. The experiments were performed on angular sand particles, as well as spherical glass beads, sheared in a cylindrical torsional rheometer with parallel plate geometry. They observe pronounced, reversible non-monotonic variation of shear band thickness as a function of shear rate in angular sand, but not in glass beads. Specifically, in the absence of tapping, the angular sand shear band thickness approaches a constant value at very slow shear rates, then dips by a maximum of roughly $10\%$ at intermediate shear rates, before increasing again at fast shear rates. The shear band thickness also varies non-monotonically with the shear rate in the presence of tapping; it is smaller than in the absence of tapping in the slow, quasistatic regime, but coincides with the no-tapping behavior in the fast, inertial regime. This is in addition to the slow compaction of the non-shearing bulk, which is not shown in the figure, and may be interpreted as aging in the presence of gravity.

\begin{figure}[here]
\centering 
\includegraphics[width=.45\textwidth]{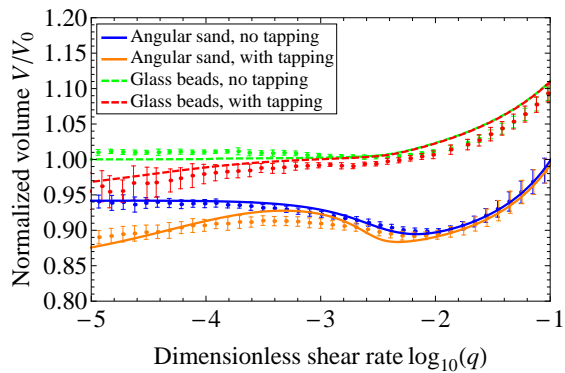} 
\caption{\label{fig:Vq}(Color online) Variation of steady-state volume $V$ or shear band thickness, normalized by the volume $V_0 = N a^3$ of grains, as a function of the dimensionless shear rate $q = \tau \dot{\gamma}^{\text{pl}}$. Results are shown for both angular sand particles and spherical glass beads. The data points indicate the average measurement over experimental runs (16 for angular sand without tapping, 30 for angular sand with tapping, 17 for glass beads without tapping, and 6 for glass beads with tapping). The error bars indicate the standard deviation of the measurements. The curves show results of the calculations using our theoretical model; parameter values used to compute these curves are summarized in Table~\ref{tab:parameters}.} 
\end{figure}

In computing the theoretical curves in Fig.~\ref{fig:Vq} we used Eqs.~\eqref{eq:X_ss2} and \eqref{eq:V4} for the steady-state volume, but drop the $\eta$-dependent term in Eq.~\eqref{eq:V4} for spherical glass beads for which the misalignments in angular grains have no counterpart. We also made a number of simplifications, and appealed to the observations to determine a number of elements in our theory. First, we neglect the diffusion of configurational disorder across the shear band boundary, and neglect aging effects in the non-shearing bulk. This assumption is justified as long as the initial state is one with a small degree of configurational disorder, for which prior STZ analyzes indeed predict the emergence of a disorder-limited shear band that relaxes very slowly if at all, with a sharp shear band boundary~\cite{lieou_2014,manning_2007a,manning_2009,daub_2009}, and within which the distribution of configurational disorder is uniform. Thus, we confine the subsequent analysis to within the shear band, and assume that internal state variables carry no spatial dependence.

Then we use the shear band thickness in the fast and slow shear rate limits, and in the absence of tapping ($\rho = 0$), to constrain the angular sand frictional noise strength $\xi$ which first appeared in Eq.~\eqref{eq:A}. We argued in Sec.~II above that $\chi^{\text{ss}} \rightarrow \hat{\chi}(q = 0) = \hat{\chi}_0$ in the limit of vanishingly slow shear rate, and that $\chi^{\text{ss}} \rightarrow \hat{\chi} (q)$ and diverges at large shear rate, but that $\theta < \chi^{\text{ss}} < \hat{\chi}(q)$ between the two limits, implying that it is plausible for $\xi (\Gamma) \sim \Gamma^2$ at small $\Gamma$, and saturates at large $\Gamma$. One way to interpolate between the slow, quasistatic and fast, inertial behaviors is to assume that $\xi$ takes the form
\begin{equation}\label{eq:xi}
 \xi (\Gamma) = \xi_0 \tanh ( \beta \Gamma^2 ) .
\end{equation}
In the ensuing analysis, we use $\xi_0 = 1.2$ and $\beta = 20$ for angular sand. On the other hand, we set $\xi = 0$ for spherical glass beads. In fact, the data points in Fig.~\ref{fig:Vq} for spherical glass beads sheared in the absence of tapping indicate a small degree of compaction at $q \sim 10^{-2}$, suggestive of a small, nonzero $\xi$. Thus, setting $\xi = 0$ for glass beads results in a small misfit between theory and data, which is not surprising because there is nonzero interparticle friction even between spherical glass beads. However, we do not attempt to model that behavior.

Van der Elst \textit{et al.}~measured the shear rate in terms of the rotation angular velocity $\omega$ of the shear cell. (The geometry is equivalent to that of a rectangular shear cell with periodic boundary conditions in the shearing direction, at least locally.) To convert this to our shear rate, we use the estimate $\dot{\gamma}^{\text{pl}} = r \omega / h$, where $r$ is the shear cell radius and $h$ is the shear band height. In their experiments, $r = 9.5$ mm and $h \simeq 1$ mm. Next, the angular sand particles have a typical diameter of $a = 350$ $\mu$m, with mass density $\rho_G = 1600$ kg m$^{-3}$, and the experiments were performed at a confining pressure of $p \simeq 7$ kPa. This gives the inertial time scale $\tau = a \sqrt{\rho_G / p} = 1.67 \times 10^{-4}$ s, so that the conversion formula between the rotation velocity $\omega$ and our dimensionless shear rate $q$ is
\begin{equation}
 q = (1.59 \times 10^{-3} \text{s}) \omega .
\end{equation}
We note in passing that the tapping frequency in the experiments is 40.2 kHz, the inverse of which is only an order of magnitude smaller than the inertial time scale. Recall our argument in Eq.~\eqref{eq:eta} that misalignments are created and annihilated by noise; thus the fact that the inertial and tapping time scales are comparable justifies our assumption that both the STZ density $\Lambda$ and the misalignment bias $\eta$ are functions of the same compactivity $\chi$. This need not be the case if the two time scales are several orders of magnitude apart; we will comment on its implication in Sec.~V.

The stress measurements were too noisy for us to be able to constrain parameters associated with STZ dynamics. However, based on other shearing experiments on angular grains~\cite{mair_1999,mair_2002}, the yield stress should be about 0.4 times the confining pressure, so we have chosen $s_0 = 0.4 p$. We have also chosen ${\cal C}(s) \simeq {\cal C}_0 = 1$ to be a constant, because the STZ transition rate should not be very sensitive to the shear stress to pressure ratio provided that $s / p < 1$. The parameter fitting procedure reveals that the steady-state volume variation is insensitive to these STZ dynamics parameters in comparison with those involved in the choice of $\hat{\chi} (q)$. We choose $\theta = 0.2$, $\rho = 0$ and $\theta = 0.18$, $\rho = 5 \times 10^{-4}$ in the absence and presence of tapping, respectively. With $\theta$ subsuming all kinetic degrees of freedom, its different values in the two cases reflect the expectation that tapping can increase the packing fraction of a generic granular assembly~\cite{daniels_2005,daniels_2006,nowak_1998,knight_1995}. (Following~\cite{edwards_1998}, it might be possible to determine $\theta$ using the fluctuation-dissipation theorem and the Langevin equation, but that gives us another adjustable parameter interpreted as a drag coefficient, so we regard $\theta$ as an adjustable parameter itself.) The other parameters in our calculation are summarized in Table~\ref{tab:parameters}. While our model produces qualitative agreement with the experiment over a wide range of parameter values, the parameter values used in the present analysis have been chosen to provide quantitative fit with the experimental measurements and, based on past experience, are physically reasonable. It may be possible to further constrain the parameters of microscopic origin with additional simple experiments; for example, stress parameters may be constrained using slide-hold-slide experiments.

\begin{table*}[t]
\caption{\label{tab:parameters}List of parameter values in our theoretical model that describes the van der Elst \textit{et al.}~experiments. Most of these parameters are of microscopic origin and have no implication on the qualitative aspect of compactivity, or effective temperature, dynamics.}
\begin{center}
\begin{tabular}{ >{\centering\arraybackslash}m{.08\textwidth} >{\centering\arraybackslash}m{.50\textwidth} >{\centering\arraybackslash}m{.11\textwidth} >{\centering\arraybackslash}m{.27\textwidth}}
\hline
Parameter & Description & Value & Reference or remark \\
\hline
$p$ & Confining pressure & 7 kPa & Constrained by experiment~\cite{brodsky_2012} \\
$s_0$ & Minimum flow stress & 2.8 kPa & Determined empirically~\cite{mair_1999,mair_2002} \\
$\tau$ & Inertial time scale & $ 1.67 \times 10^{-4}$ s & Constrained by experiment~\cite{brodsky_2012} \\
${\cal C}_0$ & Characteristic STZ transition rate & 1 & Microscopic~\cite{lieou_2012} \\
$\xi_0$ & Maximum frictional noise strength & 1.2 & Adjustable parameter \\
$\beta$ & Parameter in frictional noise strength, Eq.~\eqref{eq:xi} & 20 & Adjustable parameter \\
$\rho$ & Tapping intensity & 0, $5 \times 10^{-4}$ & Adjustable parameter \\
$\theta$ & Kinetic temperature & 0.2, 0.18 & Adjustable parameter \\
$\hat{\chi}_0$ & Steady-state dimensionless compactivity in $q \rightarrow 0$ limit & 0.3 & Adjustable parameter \\
$\hat{\chi}_1$ & Parameter in VFT expression, Eq.~\eqref{eq:vft} & 0.02 & \cite{lieou_2012} \\
$\hat{\chi}_A$ & Parameter in VFT expression, Eq.~\eqref{eq:vft} & 0.33 & \cite{lieou_2012} \\
$A$ & Parameter in VFT expression, Eq.~\eqref{eq:vft} & 2 & \cite{lieou_2012} \\
$q_0$ & Critical strain rate & 2 & \cite{lieou_2012} \\
$\epsilon_0$ & Plastic core volume per STZ in units of grain volume & 1.5 & Microscopic~\cite{lieou_2012} \\
$\epsilon_Z$ & Excess volume per STZ in units of grain volume & 0.5 & Determined Microscopic~\cite{lieou_2012} \\
$\epsilon_1$ & Effective volume expansion coefficient & 0.3 & Microscopic \\
$\epsilon_a$ & Misalignment defect volume in units of grain volume & 0.1 & Microscopic \\
\hline
\end{tabular}
\end{center}
\end{table*}

\begin{figure}[here]
\centering 
\includegraphics[width=.45\textwidth]{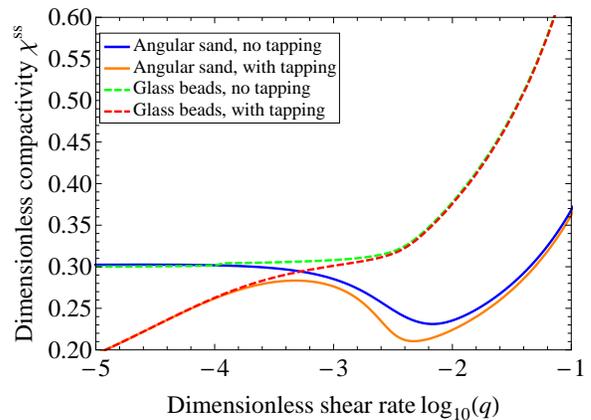} 
\caption{\label{fig:Xq}(Color online) Theoretical results for the variation of steady-state compactivity $\chi^{\text{ss}}$ with the dimensionless shear rate $q = \tau \dot{\gamma}^{\text{pl}}$, for both angular sand particles and spherical glass beads. Parameter values for each of these curves are summarized in Table~\ref{tab:parameters}.} 
\end{figure}

\begin{figure}[here]
\centering 
\includegraphics[width=.45\textwidth]{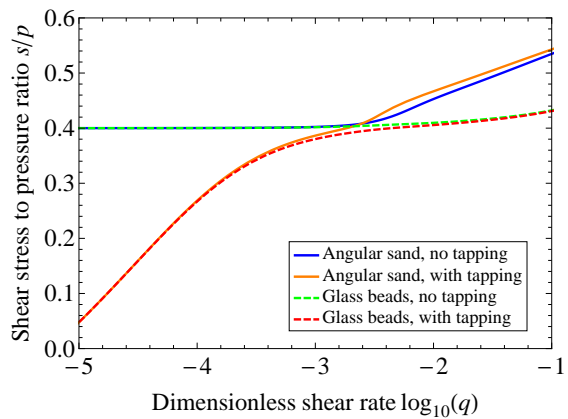} 
\caption{\label{fig:sq}(Color online) Theoretical results for the variation of shear stress to pressure ratio $s/p$ with the dimensionless shear rate $q = \tau \dot{\gamma}^{\text{pl}}$, for both angular sand particles and spherical glass beads. Parameter values for each of these curves are summarized in Table~\ref{tab:parameters}.} 
\end{figure}

In Fig.~\ref{fig:Xq}, we plot the variation of the dimensionless steady-state compactivity $\chi^{\text{ss}}$ with the shear rate $q$. $\chi^{\text{ss}}$ varies in the same qualitative manner as the volume $V$, as it should, for the two quantities are related to each other in a monotonic fashion according to Eq.~\eqref{eq:V4}. Finally, in Fig.~\ref{fig:sq}, we show our model prediction for the variation of the shear stress to pressure ratio $s / p$ with the shear rate $q$. Obviously, in the absence of vibrations, the granular medium does not flow until the shear stress exceeds the threshold $s_0$. On the other hand, tapping un-jams the system and causes it to flow at arbitrarily small shear stress $s$. This is a hallmark of glassy behavior, as seen in other amorphous solids~\cite{lieou_2012,liu_2011,haxton_2012,langer_2008}. The pronounced weakening is also hypothesized as a consequence of acoustic fluidization in earthquake faults~\cite{melosh_1979,melosh_1996}. In addition, the shear stress $s$ increases faster with the shear rate $q$ when $q > 10^{-3}$ in angular sand than in glass beads, conforming with the intuition that more work is necessary to cause angular, frictional particles to flow under shear.

\section{Concluding remarks}

In this paper, we are proposing a fundamentally new, thermodynamic interpretation of an unexpected experimental observation -- the non-monotonic variation of steady-state shear band thickness, or sample volume, as a function of shear rate, in a granular medium composed of angular, frictional particles. In our theory, this volume is determined by the effective disorder temperature (or ``compactivity'') of the grains. As a macroscopic state variable, the compactivity is controlled by a variety of microscopic mechanisms including STZ transitions, intergranular friction, and the strength of noisy fluctuations generated by collisions and external tapping. For example, energy dissipated by friction effectively ``cools'' the system and reduces its volume. The various microscopic mechanisms are described by physical parameters whose orders of magnitude can be estimated empirically and from experimental data. We have argued that if the frictional ``noise strength'' $\xi$ varies quadratically with the shear rate at the quasistatic shear limit -- in conformity with the energy dissipation associated with inelastic collisions between particles -- and saturates in the inertial, fast-shear limit, then it is possible for the steady-state compactivity and therefore the volume to show non-monotonic variation with shear rate. In other words, energy dissipated by friction effectively ``cools'' the system and reduces its volume.

In addition, we have introduced a microscopic model that quantifies the volume of the granular assembly $V$ as a function of the compactivity $\chi$. The most salient feature of this model is the combination of STZ's, soft spots with excess free volume that facilitate grain rearrangement, and misalignment defects, arising from grain interlocking and geometric frustration ubiquitous to angular particles. With a judicious choice of parameters, we have shown excellent quantitative agreement between our theory and the experimental measurements of van der Elst \textit{et al.}~\cite{brodsky_2012} on sheared angular sand particles and spherical glass beads. In our opinion, the fact that this picture fits together as well as it does is strong evidence in favor of its validity. This is not pure phenomenology. It is a systematic attempt to develop a first-principles theory of a complex and important class of physical phenomena.

We emphasize again that qualitatively, the non-monotonic variation of sample thickness with shear rate is a consequence of frictional noise alone, and not uniquely described by our microscopic model of STZ's and misalignments. One example of an alternative, simple model is the linear model $V = C \chi + D$ without the misalignment term, where the effective volume expansion coefficient $C$ is a constant. However, we have not been able to fit quantitatively the experimental data with this model nearly as closely as with the model of misalignments, which amplifies the amount of compaction at intermediate strain rates. Therefore, we have chosen to adopt the present model of STZ's and misalignments, the latter of which is necessary to account for the rather substantial amount of compaction observed in the transitional regime, between the slow, quasistatic and fast, inertial limits.

A key feature of the Ising-like model of misalignments is the existence of a ``ferromagnetic'' transition (see Eq.~\eqref{eq:eta_eq} above); it happens above a critical compactivity $\chi_c$ at which the volume variation as a function of shear rate ought to display a small cusp, at a shear rate apparently not probed by the van der Elst \textit{et al.}~experiments. Our binary, Ising-like model of misalignments might also be useful in formulating a description of other glassy phenomena in granular materials, in a manner similar to binary clusters introduced in \cite{langer_2013}.

A central assumption in the paper is that both misalignments and STZ's are governed by the same ``configurational temperature'' or compactivity $\chi$, the justification of which is that the inertial time scale $\tau = a \sqrt{\rho_G / p}$ and the inverse tapping frequency differ by less than an order of magnitude. Had this not been the case, it might be necessary to characterize the system with as many as three temperature-like quantities: the kinetic temperature $\theta$, the ``noise'' compactivity $\chi_K$ that pertains to the vibrational subsystem and governs the misalignment bias $\eta$, and the configurational compactivity $\chi_C$ that pertains to the shearing, configurational subsystem and governs the STZ density $\Lambda$, all falling out of equilibrium with one another. When that happens, the variation of volume $V$ as a function of shear rate $q = \tau \dot{\gamma}^{\text{pl}}$ in the presence of tapping need not coincide with that in the absence of tapping, in the fast shear rate limit. There are multiple ways to separate the vibrational and configurational time scales, and verify or dismiss our speculation. For example, one could conduct the shear experiment at substantially higher confining pressure $p$ to shorten the inertial time scale $\tau$ of the configurational subsystem, or tap the system at a higher frequency so that the inertial and vibrational time scales are at least several orders of magnitude apart. In the former case, we speculate that the vibrational subsystem would fall out of equilibrium with the configurational subsystem, with $\chi_K > \chi_C$; in the latter case, $\chi_C > \chi_K$. Either way, the coupling between the different subsystems would differ from that in the present paper (cf.~Eq.~\eqref{eq:X_ss2}), and qualitatively distinct behaviors might emerge. Such experiments might provide the most stringent tests yet of non-equilibrium thermodynamics.

\section*{Acknowledgments}

We thank Nicholas van der Elst and Emily Brodsky for sharing their data with us, and for illuminating discussions. Additionally, we thank Kenneth Kamrin, Bulbul Chakraborty, Karen Daniels, Karin Dahmen, and Michael Cates for sharing their insights. This work was supported by an Office of Naval Research MURI Grant No. N000140810747, NSF Grant No. DMR0606092, and the NSF/USGS Southern
California Earthquake Center, funded by NSF Cooperative Agreement EAR-0529922 and USGS Cooperative Agreement 07HQAG0008, and the David and Lucile Packard Foundation. Additionally, JSL was supported in part by the U.S. Department of Energy, Office of Basic Energy Sciences, Materials Science and Engineering Division, DE-AC05-00OR-22725, through a subcontract from Oak Ridge National Laboratory.

\end{document}